\documentclass[%
 reprint,
superscriptaddress,
 amsmath,amssymb,
 aps,
 prl,
]{revtex4-2}

\usepackage{ulem}
\usepackage{soul}
\usepackage{gensymb}
\usepackage{textcomp}
\usepackage{hyperref}
\usepackage{graphicx}
\usepackage{dcolumn}
\usepackage{bm}
\usepackage{xcolor}

\begin{document}

\preprint{APS/123-QED}

\title{Charge density wave in kagome lattice intermetallic ScV$_6$Sn$_6$ }

\author{Hasitha W. Suriya Arachchige}
\email{ssuriyaa@vols.utk.edu}
\affiliation{Department of Physics \& Astronomy, University of Tennessee Knoxville, Knoxville, Tennessee 37996, USA}%

\author{William R. Meier}
\email{javamocham@gmail.com}
\affiliation{Materials Science \& Engineering Department, University of Tennessee Knoxville, Knoxville, Tennessee 37996, USA}%

\author{Madalynn Marshall}
\affiliation{Neutron Scattering Division, Oak Ridge National Laboratory, Oak Ridge, Tennessee 37831, USA}%

\author{Takahiro Matsuoka}
\affiliation{Materials Science \& Engineering Department, University of Tennessee Knoxville, Knoxville, Tennessee 37996, USA}%

\author{Rui Xue}
\affiliation{Department of Physics \& Astronomy, University of Tennessee Knoxville, Knoxville, Tennessee 37996, USA}%

\author{Michael A. McGuire}
\affiliation{Materials Science \& Technology Division, Oak Ridge National Laboratory, Oak Ridge, Tennessee 37831, USA}%
\author{Raphael P. Hermann}
\affiliation{Materials Science \& Technology Division, Oak Ridge National Laboratory, Oak Ridge, Tennessee 37831, USA}%

\author{Huibo Cao}
\affiliation{Neutron Scattering Division, Oak Ridge National Laboratory, Oak Ridge, Tennessee 37831, USA}%

\author{David Mandrus}
\email{dmandrus@utk.edu}
\affiliation{Department of Physics \& Astronomy, University of Tennessee Knoxville, Knoxville, Tennessee 37996, USA}%
\affiliation{Materials Science \& Engineering Department, University of Tennessee Knoxville, Knoxville, Tennessee 37996, USA}%
\affiliation{Materials Science \& Technology Division, Oak Ridge National Laboratory, Oak Ridge, Tennessee 37831, USA}%

\date{\today}

\begin{abstract}
Materials hosting kagome lattices have drawn interest for the diverse magnetic and electronic states generated by geometric frustration. In the $A$V$_3$Sb$_5$ compounds ($A$ = K, Rb, Cs), stacked vanadium kagome layers give rise to unusual charge density waves (CDW) and superconductivity.
Here we report single-crystal growth and characterization of ScV$_6$Sn$_6$, a hexagonal HfFe$_6$Ge$_6$-type compound that shares this structural motif. We identify a first-order phase transition at 92\,K. Single crystal X-ray and neutron diffraction reveal a charge density wave modulation of the atomic lattice below this temperature. This is a distinctly different structural mode than that observed in the $A$V$_3$Sb$_5$ compounds, but both modes have been anticipated in kagome metals. The diverse HfFe$_6$Ge$_6$ family offers more opportunities to tune ScV$_6$Sn$_6$ and explore density wave order in kagome lattice materials.\footnote{This manuscript has been authored by UT-Battelle, LLC under Contract No. DE-AC05-00OR22725 with the U.S. Department of Energy.  The United States Government retains and the publisher, by accepting the article for publication, acknowledges that the United States Government retains a non-exclusive, paid-up, irrevocable, world-wide license to publish or reproduce the published form of this manuscript, or allow others to do so, for United States Government purposes.  The Department of Energy will provide public access to these results of federally sponsored research in accordance with the DOE Public Access Plan (http://energy.gov/downloads/doe-public-access-plan).}

\end{abstract}

\maketitle


\label{sec:Intro}

    \begin{figure}
    \includegraphics[width=8.6cm]{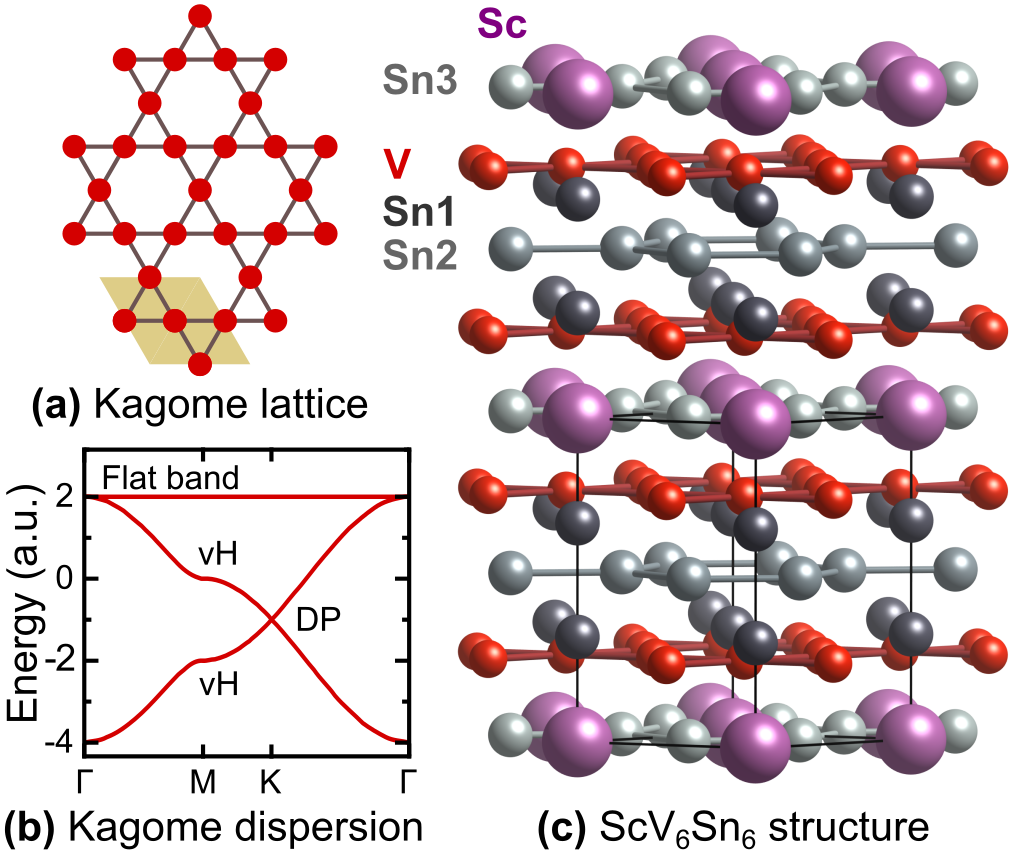}
    \caption{\label{fig:Intro} 
    	\textbf{(a)} kagome lattice \textbf{(b)} tight-binding kagome lattice electron dispersion from Ref.~\cite{Beugeling2012_TopoPhaseTransitionNNNHoppingIn2DLattices} highlighting flat band, van Hove points (vH), and Dirac points (DP). \textbf{(c)} ScV$_6$Sn$_6$ structure generated with Vesta \cite{Momma2011_Vesta3}.
    }	
    \end{figure}

	\label{sec:Intro_kagome}
    Geometrically frustrated atomic arrangements hold special interest for condensed matter physics. One example, the kagome lattice, is composed of triangles and hexagons (Fig.~\ref{fig:Intro}\textbf{(a)}). Tight binding models yield an electronic structure hosting Dirac nodes, van Hove singularities and flat bands (Fig.~\ref{fig:Intro}\textbf{(b)})\cite{Guo2009_TopoInsulatorKagomeLattice,Beugeling2012_TopoPhaseTransitionNNNHoppingIn2DLattices}. Depending on band filling and interactions, a wide variety of electronic states can be realized including charge density waves (CDWs), spin density waves, bond order, and superconductivity \cite{Christensen2021_TheoryCDW-AV3Sb5,Wang2013_CompetingElectronicOrdersKagomeLatticeAtVanHove,Tan2021_CDW+ElectronicPropsSuperconductinKagomeMetals,Feng2021_ChiralFluxPhaseKagomeSuperconductorAV3Sb5,Isakov2006_HardCoreBosonsOnKagome-ValanceBondSolids,OBrien2010_StronglyCorrelatedFermionsOnKagomeLattice,Rueegg2011_FractionallyChargedTopologicalPointDefectsOnKagome,Wen2010_InteractionDrivenTIOnKagome+DecHoneycomb,Nishimoto2010_MetalInsulatorTransOnThirdFilledKagome,Kiesel2013_FermiSurfaceInstabilitiesKagomeHubbardModel,Barros2014_ExoticMagneticOrderingsInKagomeKondoLatticeModel,Yu2012_ChiralSuperconductivity+ChiralSpinDensityWaveInHubbardKagomeModel,Denner2021_AnalysisOfChargeOrderAV3Sb5,Feng2021_LowEnergyTheory+SymmetryClassificationOfFluxPhasesOnKagome,Kiesel2012_SublatticeInterferenceInKagomeHubbardModel,Ferrari2022_CDWinKagomeExtendedHubbardAtVanHoveFilling,Ko2009_DopedKagomeSystemAsExoticSuperconductor,Park2021_ElectronicInstabilitesInKagomeMetals}.

	\label{sec:Intro_AV3Sb5}
    The hexagonal $A$V$_3$Sb$_5$ compounds ($A$ = K, Rb, Cs) are a rich family of strongly correlated materials \cite{Ortiz2019_NewKagomePrototypeMaterials-DiscoveryAV3Sb5,Ortiz2021_SuperconductivityInZ2KagomeMetalKV3Sb5,Yin2021_Superconductivity+NormStatePropertiesRbV3Sb5Crystals,Ortiz2020_CsV3Sb5-Z2TopoKagomeMetalWithSuperconductingGrdState}. Stacked vanadium kagome layers in these materials give rise to 
    CDWs and superconductivity. The origin of the CDW is closely tied to van Hove singularities at the Fermi level \cite{Liang2021_3D-CDW+SurfaceDependentVortexCoreStatesKagomeCsV3Sb5,Kang2022_vanHoveSingularities+OriginChargeOrderInCsV3Sb5,Ortiz2021_FermiSurfMapping+NatureOfCDWInCsV3Sb5,Hu2022_NatureOfVanHoveSingularitiesCsV3Sb5}. The unusual characteristics of the CDW \cite{Ortiz2021_SuperconductivityInZ2KagomeMetalKV3Sb5,Kang2022_vanHoveSingularities+OriginChargeOrderInCsV3Sb5,Nie2022_CDW-DrivenNematicityInCsV3Sb5,Wang2021_ChiralCDW-CsV3Sb5,Li2021_CDW-WithoutAcosticPhononAnomalyRbV3Sb5+CsV3Sb5,Yu2021_AnomalousHall+CDWCsV3Sb5,Jiang2021_ChiralCDW-KV3Sb5,Yang2020_AnomalousHallInKV3Sb5} and its interaction with superconductivity \cite{Yu2021_CompetitionOfSuperconductivityAndCDWInCsV3Sb5,Chen2021_RotonPairDensityWaveInCsV3b5,Du2021_PressureInducedDoubleSuperconductingDomes+ChargeInstabilityKV3Sb5,Oey2022_FermiLevel+DoubleDomeSuperconductivitySnDopedCsV3Sb5} have sparked great interest in kagome lattice derived charge order \cite{Neupert2021_ChargeOrder+SuperconductivityAV3Sb5}.

	\label{sec:Intro_166}
	
    The hexagonal HfFe$_6$Ge$_6$-type ``166'' compounds are a large family of intermetallics related to CoSn \cite{Venturini2006_HfFe6Ge6TypeAndRelatedStructures,ghimire2020competing}. Unlike CoSn and $A$V$_3$Sb$_5$, these \textit{RM}$_6$\textit{X}$_6$ compounds have two kagome sheets per unit cell separated by alternating $RX_2$ and $X_2$ layers (Fig.~\ref{fig:Intro}\textbf{(c)}). We focus on the \textit{R}V$_6$Sn$_6$ compounds as they host the vanadium kagome lattice so integral to the exciting behavior in $A$V$_3$Sb$_5$ \cite{Pokharel2021_ElectronicPropertiesOfTopologicalYV6Sn6+GdV6Sn6,Ishikawa2021_GdV6Sn6Properties,Pokharel2022_AnisoMag-TbV6Sn6,Rosenberg2022_FerromagTbV6Sn6,Romaka2019_Lu-V-Ge_Lu-V-SnSystems,Peng2021_KagomeBandStructure+KagomeSurfaceStatesGdV6Sn6+HoV6Sn6}. In fact, the rare earth variants even possess similar filling of the vanadium $d$-orbital bands \cite{Ortiz2019_NewKagomePrototypeMaterials-DiscoveryAV3Sb5,Pokharel2021_ElectronicPropertiesOfTopologicalYV6Sn6+GdV6Sn6,Hu2022_NatureOfVanHoveSingularitiesCsV3Sb5,Rosenberg2022_FerromagTbV6Sn6}. Although the band structure and $f$-orbital magnetism have garnered some interest in $R$V$_6$Sn$_6$ ($R$ = Y, Gd-Tm, and Lu) \cite{Pokharel2021_ElectronicPropertiesOfTopologicalYV6Sn6+GdV6Sn6,Hu2022_TopoSurfStates-vanHoveARPES-GdV6Sn6,Ishikawa2021_GdV6Sn6Properties,Peng2021_KagomeBandStructure+KagomeSurfaceStatesGdV6Sn6+HoV6Sn6,Pokharel2022_AnisoMag-TbV6Sn6,Rosenberg2022_FerromagTbV6Sn6,Lee2022_MagPropertiesRV6Sn6,Zhang2022_Elec+MagPropsTbV6Sn6-TmV6Sn6} no vanadium-driven order has been observed to date.
	
	In this Letter we examine the low-temperature behavior of single crystals of the kagome metal ScV$_6$Sn$_6$. Physical property measurements reveal a first-order phase transition around 92\,K reminiscent of that in the $A$V$_3$Sb$_5$ compounds. X-ray and neutron diffraction reveal a CDW	below this temperature that is distinctly different than those in $A$V$_3$Sb$_5$ \cite{Ortiz2020_CsV3Sb5-Z2TopoKagomeMetalWithSuperconductingGrdState,Liang2021_3D-CDW+SurfaceDependentVortexCoreStatesKagomeCsV3Sb5,Ortiz2021_FermiSurfMapping+NatureOfCDWInCsV3Sb5,Wang2021_ChiralCDW-CsV3Sb5,Jiang2021_ChiralCDW-KV3Sb5}. The CDW we identify in ScV$_6$Sn$_6$ demonstrates that charge order is a common feature in partly filled $d$-orbital kagome systems.  Compared to the $A$V$_3$Sb$_5$ compounds, the HfFe$_6$Ge$_6$-type compounds offer improved tuneability making them an ideal platform to explore the curious CDWs in transition metal kagome systems.

\section{Methods}
\label{sec:Methods}

\label{sec:Methods_Growth}
Single crystals were grown from a molten Sn flux using an atomic ratio of Sc:V:Sn = 1:6:60. Dendritic scandium metal (Alfa Aesar 99.9\%), vanadium pieces (Alfa Aesar 99.8\%), and Sn shot (Alfa Aesar 99.99+\%) were loaded into a 2\,ml alumina Canfield crucible set \cite{Canfield2016_FritDiskCrucibles}. The crucible assembly was sealed in an argon-filled fused silica ampule. Subsequently, the ampule was heated up to 1150\,\textdegree C over 12\,h followed by a 15\,h dwell. Crystals were grown during a slow cool to 780\,\textdegree C at 1\,\textdegree C/h. Single crystals were extracted from the melt by centrifuging.

The ScV$_{6}$Sn$_{6}$ crystals are nicely faceted metallic barrel-shaped hexagonal blocks roughly 1-3\,mm in size and their HfFe$_6$Ge$_6$-type structure was established by powder x-ray diffraction (supplemental information). 
Sn and VSn$_3$ were removed from crystals with 10 wt\% aqueous HCl.
Crystal composition was estimated using Energy Dispersive Spectroscopy (EDS) on a polished surface was Sc\,:\,V\,:\,Sn = 1\,:\,6.21(7)\,:\,6.22(11).

\label{sec:Methods_StructureCharacterization}


Low-temperature lattice parameters were estimated from Rietveld fits of powder x-ray diffraction (PXRD) patterns obtained using a Oxford PheniX closed-cycle helium cryostat on a PANalytical X’pert Pro diffactometer with a Cu $K_\alpha$ source.


Single crystal x-ray diffraction (SCXRD) studies were performed at 280\,K and 50\,K with a Rigaku XtaLAB PRO diffractometer using Mo $K_\alpha$ radiation, a Rigaku HyPix-6000HE detector and an Oxford N-HeliX cryocooler. Rigaku Oxford Diffraction CrysAlisPro \cite{CrysAlisPRO} was used for peak indexing and integration and JANA or Shelx for structural refinement \cite{Petricek2014_Jana2006}.

Single crystal neutron diffraction (SCND) was measured at HB-3A DEMAND \cite{Cao2018_DEMAND} at the High Flux Isotope Reactor at Oak Ridge National Laboratory using 1.542\,\AA~neutrons from a bent Si-220 monochromator \cite{Chakoumakos2011_HB-3A}. A $2\times2\times1$\,mm$^3$ crystal of ScV$_6$Sn$_6$ was placed inside a closed-cycle refrigerator (CCR) and the $\frac{1}{3} \frac{1}{3} \frac{10}{3}$ superlattice peak was measured on warming and cooling.

$^{119}$Sn M\"ossbauer spectra of ScV$_6$Sn$_6$ were measured at numerous temperatures on 28\,mg/cm$^2$ powder sample using a Janis SH-850 closed-cycle cryostat and a constant-acceleration Wissel drive. A 0.1 mCi $^{119\mathrm{m}}$Sn source and a 2\,in diameter Ametek Tl@NaI detector were employed with a 25\,\textmu m Pd foil between sample and detector. 

\label{sec:Methods_PhysicalProperties}
Electrical resistivity measurements were obtained using four electrodes on a bar-shaped polished crystal with current perpendicular to $\bm{c}$-axis. Resistivity was measured with a Keithley 2450 SourceMeter and a 2182 Nanovoltmeter setup controlled by LabView. Temperature control and magnetic field were provided by a Quantum Design Physical Property Measurement System (PPMS) with magnetic field along $\bm{c}$. 

For magnetization measurements, an acid-etched crystal was mounted in a plastic drinking straws in a Quantum Design MPMS 3 using the Vibrating Sample Magnetometer (VSM) option. The moment was measured between 3 and 300\,K in a field of 1\,T applied perpendicular to $\bm{c}$.

Finally, heat capacity measurements were carried out using the heat capacity option of the PPMS with a 7.95\,mg crystal mounted with Apiezon N-grease. Data were collected between 2 and 200\,K by applying a series of heat pulses that each raised the sample temperature by about 30\% and analyzing the sample temperature vs time during heating and the subsequent cooling \cite{QD-HeatCap}. Near the transition temperature, heating and cooling curves were processed separately to observe the thermal hysteresis associated with the transition. 

\section{Results}
\label{sec:Results}

\label{sec:Methods_Transition}

\begin{figure}
\includegraphics[width=8.6cm]{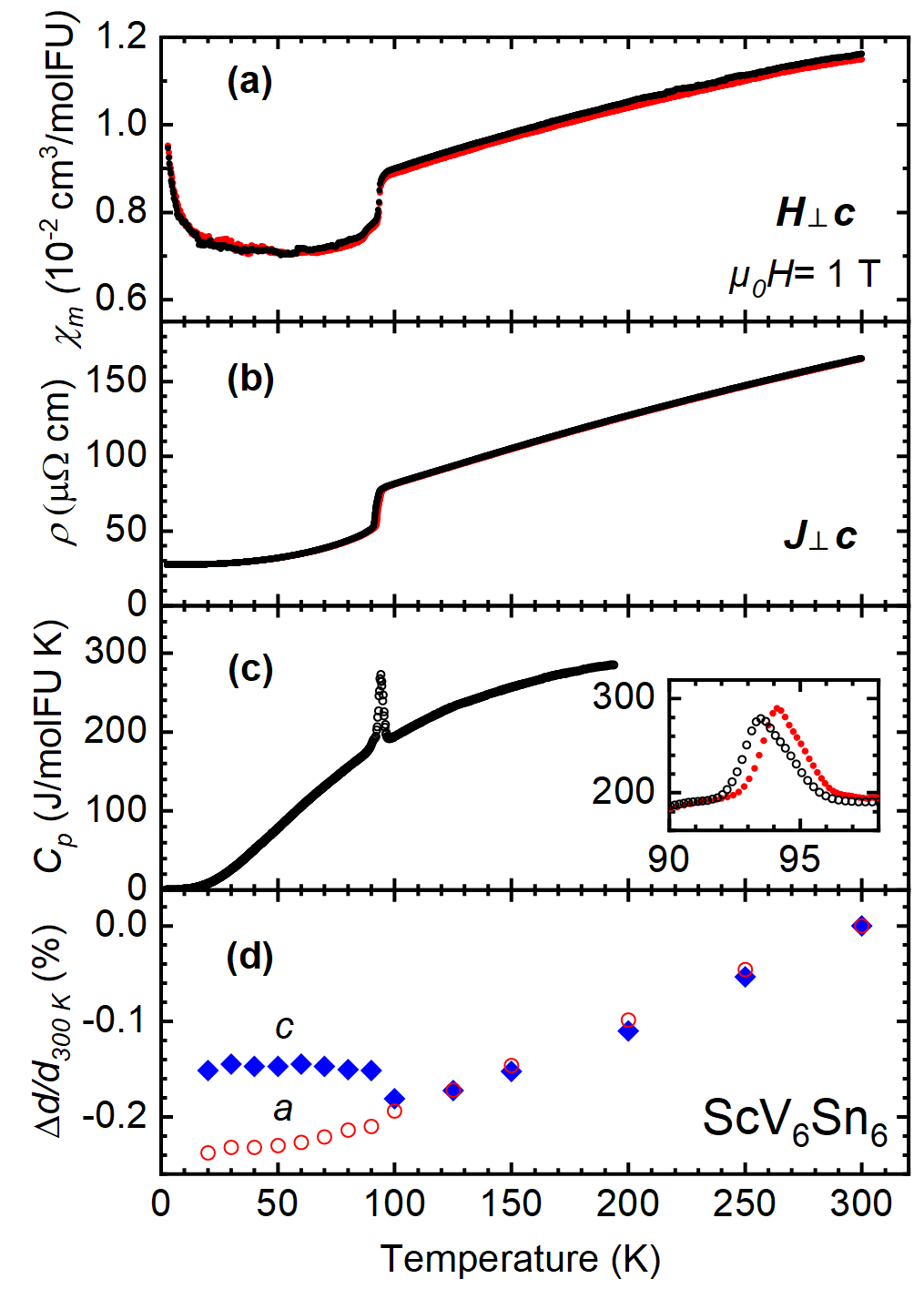}
\caption{\label{fig:PhysicalProperties}
    Low-temperature properties of ScV$_6$Sn$_6$ reveal a first-order phase transition at 92\,K. \textbf{(a)} temperature dependence of magnetic susceptibility, $\chi_m$, on cooling (black) and warming (red). \textbf{(b)} temperature dependence of resistivity, $\rho$, in the $ab$-plane measured on cooling (black) and warming (red). \textbf{(c)} Specific heat capacity of ScV$_6$Sn$_6$, $C_p$, averaging warming and cooling parts of each pulse. The inset  shows the data from the warming (red) and cooling (black) segments of each heat pulse. \textbf{(d)} Relative change in ScV$_6$Sn$_6$ lattice parameters vs. temperature.
    }
\end{figure}

Figure \ref{fig:PhysicalProperties} presents temperature-dependent measurements of ScV$_6$Sn$_6$ revealing a phase transition around 92\,K. Magnetic susceptibility ($\chi _m$) in Fig.~\ref{fig:PhysicalProperties}\textbf{(a)} reveals a weak Pauli paramagnetic response \cite{Blundell2001_MagnetismInCondensedMatter,Kittel2004_SolidStatePhysics} slightly larger than the 0.001-0.002\,cm$^3$/mol observed for YV$_6$Sn$_6$ \cite{Pokharel2021_ElectronicPropertiesOfTopologicalYV6Sn6+GdV6Sn6,Ishikawa2021_GdV6Sn6Properties}. In contrast to the yttrium compound, ScV$_6$Sn$_6$ has an abrupt 20\% drop in the susceptibility on cooling through 92\,K.

The transition observed in magnetization is corroborated by electrical resistivity, Fig.~\ref{fig:PhysicalProperties}\textbf{(b)}. ScV$_6$Sn$_6$ displays metallic resistivity decreasing from 164 to 28\,\textmu$\Omega$\,cm from 300 to 2\,K and no bulk superconductivity down to 80\,mK. Critically, there is a conspicuous 35\% drop in resistivity on cooling through 92\,K. The YV$_6$Sn$_6$, GdV$_6$Sn$_6$, and TbV$_6$Sn$_6$ do not have a step like feature and have smaller resistivities across the whole temperature range \cite{Pokharel2021_ElectronicPropertiesOfTopologicalYV6Sn6+GdV6Sn6,Ishikawa2021_GdV6Sn6Properties,Pokharel2022_AnisoMag-TbV6Sn6,Rosenberg2022_FerromagTbV6Sn6}.

Heat capacity provides clear evidence that the features in $\chi_m(T)$ and $\rho(T)$ correspond to a bulk phase transition in ScV$_6$Sn$_6$. Figure \ref{fig:PhysicalProperties}\textbf{(c)} presents the specific heat capacity, $C_p(T)$. A sharp peak in the heat capacity represents the heat of transformation across a first-order phase transition. This first-order nature of the transition is evident in the discontinuities in $\chi_m(T)$ and $\rho(T)$ as well as the $\sim$ 0.3-1\,K thermal hysteresis observed in these measurements.


The temperature dependence of the lattice parameters in Fig.~\ref{fig:PhysicalProperties}\textbf{(d)} demonstrates that the phase transition in ScV$_6$Sn$_6$ couples to the crystal lattice. On cooling through the 92\,K transition, $c$ increases by 0.04\% but $a$ remains relatively unchanged.

\begin{figure*}
\includegraphics[width=17.8cm]{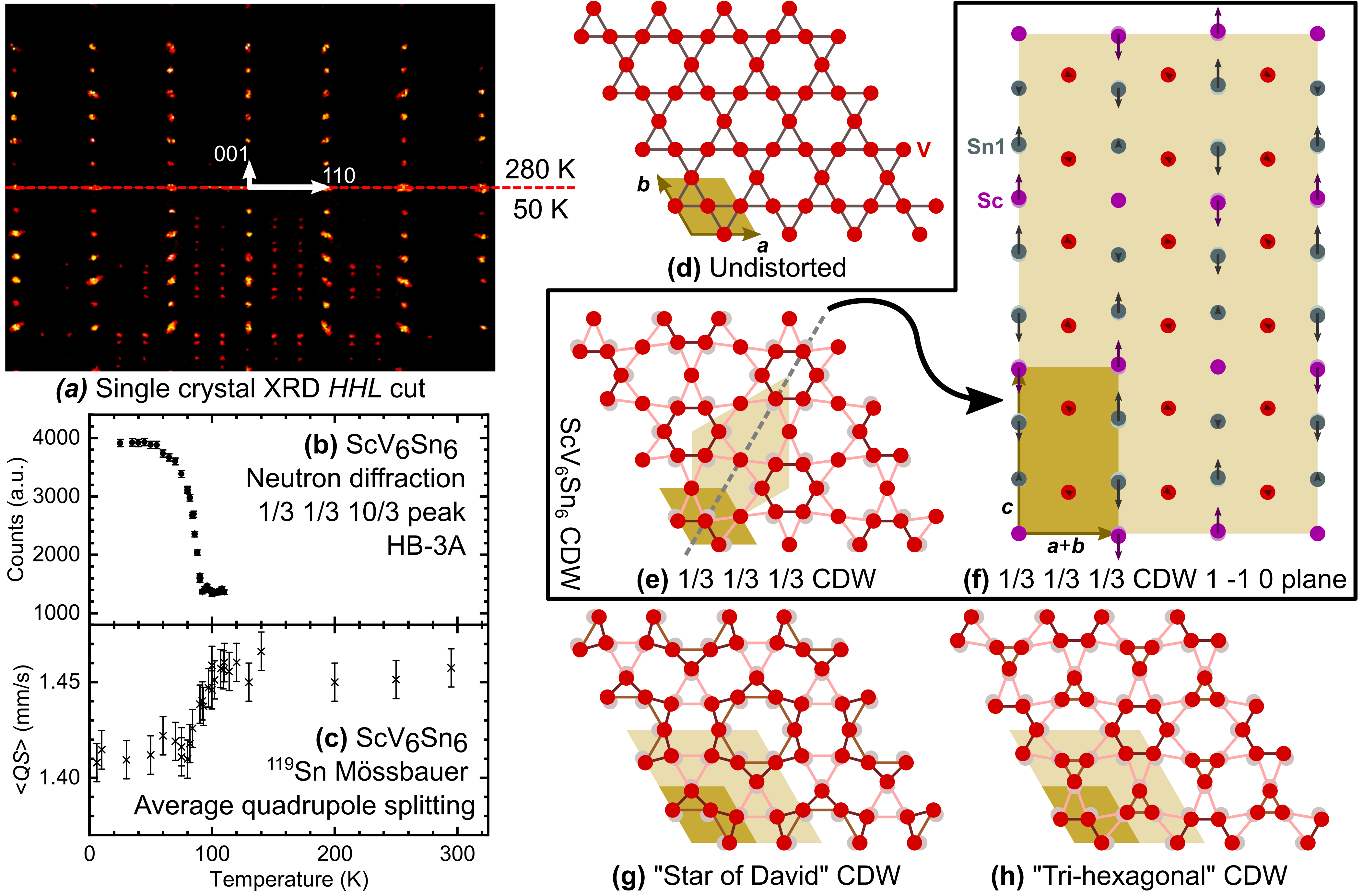}
\caption{\label{fig:CDW}
	Details of the CDW in ScV$_6$Sn$_6$ \textbf{(a)} single crystal diffraction intensity in the $HHL$ plane. New peaks at 50\,K signal a structural modulation with wave-vector $\frac{1}{3} \frac{1}{3} \frac{1}{3}$. \textbf{(b)} Single crystal neutron diffraction intensity from the $\frac{1}{3} \frac{1}{3} \frac{10}{3}$ peak.
	\textbf{(c)} Average quadrupole splitting for a three site fit to $^{119}$Sn M\"ossbauer spectra. \textbf{(d)} Kagome lattice with ScV$_6$Sn$_6$ unit cell (dark yellow). \textbf{(e)} exaggerated $P_1$ $\frac{1}{3} \frac{1}{3} \frac{1}{3}$ CDW mode showing 3-times larger in-plane unit cell (light yellow). Shorter and longer bonds are darker and lighter, respectively. \textbf{(f)} Depiction of refined atomic displacements in $\frac{1}{3} \frac{1}{3} \frac{1}{3}$ CDW mode in a (1 -1 0) plane of ScV$_6$Sn$_6$. \textbf{(g)} and \textbf{(h)} The star of David and tri-hexagonal CDW modulations of the vanadium kagome lattice in CsV$_3$Sb$_5$. These  $\frac{1}{2} \frac{1}{2} \frac{1}{2}$ or $\frac{1}{2} \frac{1}{2} \frac{1}{4}$ modes quadruple the unit cell in the $ab$-plane (light yellow rhombi).
}	
\end{figure*}

SCXRD reveals a CDW below the 92\,K phase transition in ScV$_6$Sn$_6$. Figure \ref{fig:CDW} \textbf{(a)} presents the diffracted intensity in the $HHL$ plane at 280 and 50\,K. On cooling, new spots are observed which can be indexed by a $\frac{1}{3} \frac{1}{3} \frac{1}{3}$ propagation vector. These superlattice peaks are evidence of a commensurate CDW, a periodic displacement of the atom lattice \cite{Gruner1994_DensityWavesInSolids}. SCND reveal that superlattice peak intensity appears near 92\,K and grows on cooling (Fig.~\ref{fig:CDW}\textbf{(b)}) demonstrating the connection between the CDW and the transition evident in Fig.\,\ref{fig:PhysicalProperties}.

The refined $R32$ low temperature structure is presented in supplemental materials and is dominated by the displacement mode transforming as the irreducible representation $P_1$ of $P6/mmm$ (notation from ISODISTORT and Amplimodes applications) \cite{Campbell2006_ISODISPLACE,ISODISTORT_6.7.2,Orobengoa2009_Amplimodes,PerezMato2010_Amplimodes}.

M\"ossbauer spectroscopy reveals that the Sn atoms see this structural modulation \ref{fig:CDW}\textbf{(c)}. On cooling through the 92\,K transition the absorption spectra abruptly change shape as reflected by a step of the average quadrupole splitting (Fig.\,\ref{fig:CDW}\textbf{c} and supplemental materials).

\section{Discussion}
\label{sec:Discussion}
The CDW 
in ScV$_6$Sn$_6$ has 
similarities to the CDWs in the $A$V$_3$Sb$_5$ compounds. First, both compounds host vanadium kagome lattices with V-V distances between 2.73 and 2.75\,\AA  \cite{Ortiz2019_NewKagomePrototypeMaterials-DiscoveryAV3Sb5}. The Fermi level sits within vanadium $d$-orbital bands in both the $A$V$_3$Sb$_5$ compounds \cite{Ortiz2019_NewKagomePrototypeMaterials-DiscoveryAV3Sb5,Ortiz2020_CsV3Sb5-Z2TopoKagomeMetalWithSuperconductingGrdState,Ortiz2021_FermiSurfMapping+NatureOfCDWInCsV3Sb5,Kang2022_vanHoveSingularities+OriginChargeOrderInCsV3Sb5,Neupert2021_ChargeOrder+SuperconductivityAV3Sb5,LaBollita2021_ComputationTuningVanHoveAV3Sb5ViaPressure+Doping} and GdV$_6$Sn$_6$ \cite{Pokharel2021_ElectronicPropertiesOfTopologicalYV6Sn6+GdV6Sn6,Ishikawa2021_GdV6Sn6Properties,Peng2021_KagomeBandStructure+KagomeSurfaceStatesGdV6Sn6+HoV6Sn6,Hu2022_TopoSurfStates-vanHoveARPES-GdV6Sn6}. It is likely that the vanadium bands play a key role in CDW formation in ScV$_6$Sn$_6$ as they do in the $A$V$_3$Sb$_5$ compounds \cite{Ortiz2020_CsV3Sb5-Z2TopoKagomeMetalWithSuperconductingGrdState,Kang2022_vanHoveSingularities+OriginChargeOrderInCsV3Sb5,Ortiz2021_FermiSurfMapping+NatureOfCDWInCsV3Sb5,Hu2022_NatureOfVanHoveSingularitiesCsV3Sb5}.

Second, the CDWs have a similar impact on the physical properties in these materials. For example, the CDW in CsV$_3$Sb$_5$ is accompanied by a sharp drop of both magnetic susceptibility and in-plane resistivity on cooling \cite{Ortiz2020_CsV3Sb5-Z2TopoKagomeMetalWithSuperconductingGrdState,Wang2021_ChiralCDW-CsV3Sb5,Yu2021_AnomalousHall+CDWCsV3Sb5} just as we observe in Fig.~\ref{fig:PhysicalProperties}. Discontinuities in resistance and magnetic susceptibility are frequent signatures of CDW modifications of the electronic structure
\cite{Gruner1994_DensityWavesInSolids,Meier2021_CatastrophicCDW-BaFe2Al9,Ramakrishnan2020_CDWinEr2Ir3Si5,Singh2005_CDWinLu2Ir3Si5}.

Importantly, the CDWs in $A$V$_3$Sb$_5$ family and ScV$_6$Sn$_6$ have two important differences. First, the modulation wave vectors in the two compounds are different. We observe a $\frac{1}{3} \frac{1}{3} \frac{1}{3}$ wave vector in ScV$_6$Sn$_6$ tripling the $ab$-plane area of the cell (Fig.~\ref{fig:CDW}\textbf{(d)}). Along $c$, the rhombohedral structure repeats every three unit cells (Fig.~\ref{fig:CDW}\textbf{(e)} and supplemental information). In contrast, the $\frac{1}{2} \frac{1}{2} \frac{1}{2}$ or $\frac{1}{2} \frac{1}{2} \frac{1}{4}$ CDW in the $A$V$_3$Sb$_5$ compounds quadruples the unit cell in the $ab$-plane (Fig.~\ref{fig:CDW}\textbf{(f)} and\textbf{(g)}) and doubles or quadruples the $\bm{c}$-axis \cite{Ortiz2020_CsV3Sb5-Z2TopoKagomeMetalWithSuperconductingGrdState,Liang2021_3D-CDW+SurfaceDependentVortexCoreStatesKagomeCsV3Sb5,Ortiz2021_FermiSurfMapping+NatureOfCDWInCsV3Sb5,Nie2022_CDW-DrivenNematicityInCsV3Sb5,Wang2021_ChiralCDW-CsV3Sb5,Li2021_CDW-WithoutAcosticPhononAnomalyRbV3Sb5+CsV3Sb5,Neupert2021_ChargeOrder+SuperconductivityAV3Sb5,Mu2021_Superconductivity+CDWInCsV3Sb5FromSb-NQR+V-NMR}.

The second key difference between the CDWs in these compounds is the displacements of the vanadium atoms. In CsV$_3$Sb$_5$, the vanadium atoms displace in the plane by 0.009-0.085\,\AA~forming either the Star of David or tri-hexagonal arrangement (Fig.~\ref{fig:CDW}\textbf{(f)} and\textbf{(g)}) \cite{Ortiz2021_FermiSurfMapping+NatureOfCDWInCsV3Sb5}. 
In contrast, our refinement suggests that Sc and Sn1 have the largest modulated displacements in ScV$_6$Sn$_6$. We estimate Sc and Sn1 displace up to 0.16\,\AA~along the $\bm{c}$ axis. The vanadium atoms appear to have a far weaker response, displacing only 0.004-0.006\,\AA.


CDWs and bond density waves are prominent instabilities of partly filled kagome bands. Charge order is favored by nearest neighbor Coulomb repulsion for $\frac{1}{3}$ and $\frac{2}{3}$ filled kagome bands \cite{Wang2013_CompetingElectronicOrdersKagomeLatticeAtVanHove,OBrien2010_StronglyCorrelatedFermionsOnKagomeLattice,Rueegg2011_FractionallyChargedTopologicalPointDefectsOnKagome,Wen2010_InteractionDrivenTIOnKagome+DecHoneycomb,Nishimoto2010_MetalInsulatorTransOnThirdFilledKagome,Ferrari2022_CDWinKagomeExtendedHubbardAtVanHoveFilling,Park2021_ElectronicInstabilitesInKagomeMetals}. In fact, the $\frac{1}{2} \frac{1}{2}$ and $\frac{1}{3} \frac{1}{3}$ in-plane modulation of the kagome sheet are favored by nearest neighbor Coulomb repulsion \cite{Wang2013_CompetingElectronicOrdersKagomeLatticeAtVanHove,Rueegg2011_FractionallyChargedTopologicalPointDefectsOnKagome,Wen2010_InteractionDrivenTIOnKagome+DecHoneycomb,Nishimoto2010_MetalInsulatorTransOnThirdFilledKagome}. The weak vanadium displacements we infer from our ScV$_6$Sn$_6$ diffraction data could be due to some of the exotic orders proposed in the kagome lattice involving chiral currents and bond orders\cite{Jiang2021_ChiralCDW-KV3Sb5,Nie2022_CDW-DrivenNematicityInCsV3Sb5,Feng2021_ChiralFluxPhaseKagomeSuperconductorAV3Sb5,Park2021_ElectronicInstabilitesInKagomeMetals,Nishimoto2010_MetalInsulatorTransOnThirdFilledKagome,Kiesel2013_FermiSurfaceInstabilitiesKagomeHubbardModel,Denner2021_AnalysisOfChargeOrderAV3Sb5,Feng2021_LowEnergyTheory+SymmetryClassificationOfFluxPhasesOnKagome}.

Upcoming investigations of the CDW in ScV$_6$Sn$_6$ have a distinct advantage over the $A$V$_3$Sb$_5$ compounds: tuneability. Although KV$_3$Sb$_5$, RbV$_3$Sb$_5$, CsV$_3$Sb$_5$ compounds have generated significant excitement, the family is quite small \cite{Ortiz2019_NewKagomePrototypeMaterials-DiscoveryAV3Sb5} (despite proposed variants\cite{Jiang2022_ScreeningCsV3Sb5LikeKagomeMaterials}). Doping opportunities are also limited to Ti or Nb for V \cite{Liu2021_TiDopingEvolutionOfCDW+SCInCsV3Sb5,Kato2022_Nb-DopedCsV3Sb5} and Sn for Sb \cite{Oey2022_FermiLevel+DoubleDomeSuperconductivitySnDopedCsV3Sb5,Oey2022_Sn-DopedKV3Sb5+RbV3Sb5}. In contrast, the \textit{RM}$_6$\textit{X}$_6$ compounds of the HfFe$_6$Ge$_6$ family are far more diverse \cite{Venturini2006_HfFe6Ge6TypeAndRelatedStructures} offering many different options for tuning the CDW in ScV$_6$Sn$_6$ to uncover what factors influence charge order and superconductivity in kagome metals.

\section{Conclusion}
The intriguing charge density waves (CDWs) in the $A$V$_3$Sb$_5$ compounds inspired us to investigate ScV$_6$Sn$_6$. Low-temperature physical property measurements reveal a first-order 92\,K transition. X-ray and neutron diffraction reveal a CDW modulation of the atomic lattice below this temperature. Despite the similarities between ScV$_6$Sn$_6$ and the $A$V$_3$Sb$_5$ compounds, their CDWs have different propagation vectors. Excitingly, ScV$_6$Sn$_6$ belongs to a large family of compounds offering more tuning opportunities to explore the origin of charge order in kagome lattice compounds.

\section{Acknowledgments}
\begin{acknowledgments}
	\label{sec:Acknowledgment}
	
	HWSA, WRM, TM, RX, and DM acknowledge support from the Gordon and Betty Moore Foundation’s EPiQS Initiative, Grant GBMF9069. MM and HC acknowledge  support from the US Department of Energy (DOE), Office of Science, Office of Basic Energy Sciences, Early Career Research Program Award KC0402020, under Contract No. DE-AC05-00OR22725. This research used resources at the High Flux Isotope Reactor and the Spallation Neutron Source, the DOE Office of Science User Facility operated by ORNL. MAM and RPH acknowledge support from the US Department of Energy, Office of Science, Basic Energy Sciences, Materials Science and Engineering Division (low-temperature powder X-ray diffraction, heat capacity measurements, and M\"ossbauer spectroscopy respectively). 
	
\end{acknowledgments}

%

\end{document}